\documentclass[11pt]{article}
\usepackage{cospar}
\usepackage{url}
\usepackage{natbib}

\usepackage{graphicx}
\usepackage[figuresright]{rotating}

\newcommand{\psr}{PSR~J1811--1925}
\newcommand{\GXI}{G11.2--0.3}
\newcommand{\xte}{\textit{RXTE}}

\title{Phase Coherent Timing of \psr: The Pulsar at the Heart of \GXI}

\author{ F.P. Gavriil  \address{Physics Department, McGill University, Montreal, QC H3A 2T8, Canada},
        V.M. Kaspi $^{1}$,
        and
        M.S.E. Roberts$^{1}$ \address{Department of Physics and Center for Space Research,
Massachusetts Institute of Technology, Cambridge, MA 02139, USA}}

\begin{document}

% typeset front matter   
\maketitle
\begin{abstract}

The X-ray Pulsar \psr\ in the historic supernova remnant \GXI\ has a characteristic age over 10 times the age of the remnant. This likely implies that its current spin period, $65$~ms, is close to its birth spin period. Alternatively, the pulsar may have an unusually high braking index. We report here on regular \textit{Rossi X-ray Timing Explorer}/ Proportional Counting Array (\xte/PCA) timing observations of the pulsar that were designed to measure its braking index. We provide a preliminary phase-coherent timing solution which includes a significant $\ddot{\nu}$. The braking index we measure is $\gg 3$, likely a result of conventional timing noise.  We also report on a preliminary analysis of the pulsar's unusually hard spectrum:  we determine a photon index of $\Gamma = 1.78\pm0.74$, for the pulsed component of the spectra, consistent within the uncertainties with previous \textit{ASCA} and \textit{Chandra} observations. The pulsed emission of \psr\  is seen beyond 30~keV, and the pulsations remain sinusoidal up to and beyond this energy. 

\end{abstract}

\section*{INTRODUCTION}

The determination of the ages of pulsars is of crucial 
importance to neutron star astrophysics.  
Neutron star population synthesis, birthrate,
supernova remnant associations,
and even cooling timescales (which serve  as probes of the neutron star
equation of state), all depend on accurate determinations of 
pulsar ages. 
In general, a pulsar's age is estimated by assuming
a spin-down evolution of the form
\begin{equation}
\dot\nu \propto \nu^n,
\label{eq:spindown}
\end{equation}
where $n$ is known as the {\it braking index}.  It can be determined
observationally by measuring $\ddot\nu $, since, as is easily shown,
$n=\nu\ddot{\nu}/\dot{\nu^2}$.  Integrating Equation~\ref{eq:spindown} 
over the
lifetime of the pulsar, assuming a constant braking index, results 
in the following estimate of the pulsar's age: 
\begin{equation}
\tau={P\over(n-1)\dot P} \left[1 - \left({P_0\over P}\right)^{n-1}\right],
\label{eq:age}
\end{equation}
where $P_0$ is the spin period of the pulsar at birth. It is conventional to 
assume $n=3$, appropriate for spin-down due to magnetic 
dipole radiation, and $P_0<<P$ to derive the characteristic age $\tau_c = P/
(2\dot P)$. 

Timing noise, stochastic variations in the pulse spin-down \cite[see e.g.][]{lpsc96}, make
accurate determinations of the braking index impossible for the large
majority of the over 1300 radio pulsars currently known. The five
reliably measured radio pulsar braking indices are in the range
$1.4-2.9$ \citep{kms+94,lpsc96,lps88,ckl+00, zm+01}.  This clearly signals a
significant difference between pulsar electrodynamics and a simple
dipole spinning down {\it in vaccuo}.  This is of major interest to
modelers of pulsar magnetospheres \cite[e.g.][]{mel97}.  But it is also
very important for neutron star astrophysics in general:  if the
braking indices of all pulsars are $n<3$, then the characteristic age
systematically underestimates the true age.

%\begin{figure}
%\begin{center}
%\includegraphics*[width=0.35\textwidth]{p0.eps}
%\caption{Pulsar age as a function of initial spin period. The true remnant age is 2~kyr and is indicated by the horizontal line. For the pulsar and the remnant ages to agree, we must have $n \gg 3$. \label{fig:braking}}
%\end{center}
%\end{figure}

On the other hand, initial spin periods $P_0$ that are a non-negligible
fraction of the current spin period $P$ lead to a systematic
over estimation of the true age. The distribution of initial spin
periods is also of intrinsic scientific interest, since it gives clues
to the physics of the formation of neutron stars as well as such
possibilities as early epochs of rapid spin-down due to gravitational
radiation or interactions with fossil disks.  However, to infer the
initial spin period from Equation~\ref{eq:age} requires an independent
determination of the true age, such as an associated historical event
or a well determined age from an associated supernova remnant. This
situation is very rare. The best, and until recently only, example is
the Crab pulsar, whose measured braking index and known age results in
an inferred $P_0\sim 19$~ms. There are very few other historical
supernovae that might be used. Of course an isolated pulsar's current
spin period is an upper limit on the initial spin period, so, for
example, $P_0 < 16$~ms can be trivially inferred  for PSR~J0537$-$6910 in the LMC \citep{mgz+98}.

\subsection{\psr\ -- The Pulsar at the Heart of \GXI}

\GXI\ is a young supernova remnant plausibly associated with the
historical ``guest star'' witnessed by Chinese astronomers in the year
386 A.D. \citep{cs77a}.  Radio and X-ray estimates of its
age \citep{ggts88,vad+96}, assuming a type II
supernova, support the historical association. A young 65-ms pulsar with spin-down energy $\dot{E} = 6.4\times10^{36}$~erg~s$^{-1}$ was
discovered in X-rays by \textit{ASCA} within G11.2$-$0.3 \citep{ttd+97}. Recent \textit{Chandra} observations place the pulsar at the precise center of 
the remnant \citep{krv+01}, making the association of the two unambiguous.
Surprisingly, the $\dot P=4.4\times 10^{-14}$, determined from one
observation in 1994 and three observations in 1998, implies a
characteristic spin-down age of $\tau_c=24,000$~yr \citep{ttd+99}, 15 times
that inferred from the historical association, the remnant radio surface
brightness, and the remnant X-ray spectrum. 
For this large discrepancy to be resolved,
either the braking index $n > 30$, or the initial spin period $P_0\sim62$~ms 
is very near the current spin period. 
Very recently, a similar $P_0$ was inferred for the newly
discovered pulsar in 3C58/SN1181 \citep{mss+02}, suggesting that
such long initial spin periods may be common. Unfortunately, the number
of systems where the initial spin period  can be estimated is likely to remain small.
Therefore, it is crucial that all alternative explanations for the age
discrepancy observed in \GXI\  be fully explored.

\section*{OBSERVATIONS}

The results presented here were obtained using the Proportional Counter Array (PCA) 
on board the \textit{Rossi X-ray Timing Explorer} (\xte). Our observations consist primarily of $\sim 20$~ks observations taken on a monthly basis.

\subsection{Phase Coherent Timing}
\label{sec:timing}

Deep radio searches have been unable to detect the pulsar
\citep{ckm+98a}, and so X-ray monitoring of the source is the
only option for timing.
The spin-down rate $\dot{P}$ of \psr\  was first made by \citet{ttd+99}
 using \textit{ASCA}, essentially on the basis of periods at 
just two widely spaced epochs.
Here we present a  phase-coherent timing solution for the the
pulsar using  regular \xte\ observations.

\begin{table}
\begin{center}
\begin{tabular}{lc}\hline\hline
%\multicolumn{2}{c}{Spin Parameters for \psr}\\
Parameter & Value \\
\hline
R.A. (J2000)  &  18$^{\rm h}$~11$^{\rm m}$~29$^{\rm s}$.22 \\
Decl. (J2000) & $-19^{\circ} \; 25' \;27.''6$  \\
%Distance (kpc)& 5\\
%\hline
Range of MJDs & 52341--52541\\
No. Pulse Arrival Times & 18\\
$\nu$ (Hz) & 15.461208393(11)\\
$\dot{\nu}$ ($\times 10^{-11}$ Hz s$^{-1}$) &  $-$1.00975(28) \\
$\ddot{\nu}$ ($\times 10^{-20}$ Hz s$^{-2}$) & 1.40(3) \\
$P$ (ms)& 64.67799741(44) \\
$\dot{P}$  ($\times 10^{-15}$ s s$^{-1}$)& 42.240(12)  \\
Epoch (MJD) & 52341.8046 \\
RMS Residual (ms) & 2.7 \\
%\hline
Surface Magnetic Field, $B_s$ ($\times 10^{12}$ G) & 1.7 \\
Spin-down Luminosity, $\dot{E}$  ($\times 10^{36}$ erg s$^{-1}$) & 6.1\\
Characteristic Age, $\tau_c$ (kyr) & 24 \\
\hline
\end{tabular}
\end{center}
\caption{Spin parameters for \psr\ derived from phase coherent timing of this source for $\sim9$~months.\label{ta:spin}}
\end{table}

\begin{figure}
\begin{minipage}{0.45\textwidth}
\begin{center}
\includegraphics*[width=0.75\textwidth]{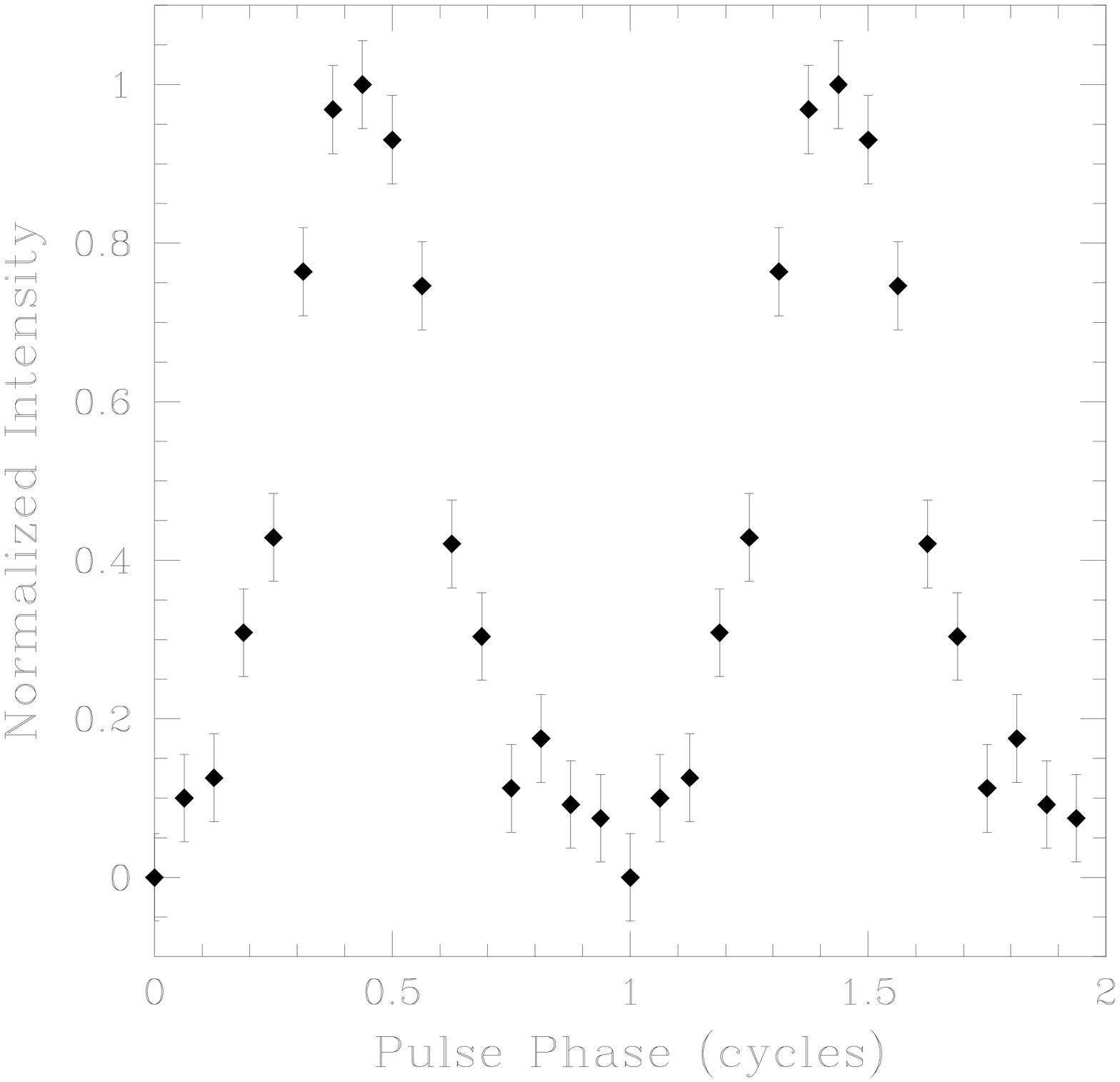}
\caption{X-ray pulse profile in the 2-30~keV range. Two cycles are shown for clarity. Profiles in the 2-10~keV and 
10-30~keV range are identical.\label{fig:profile}}
\end{center}
\end{minipage}\hfill
\begin{minipage}{0.45\textwidth}
\begin{center}
\includegraphics*[width=0.75\textwidth]{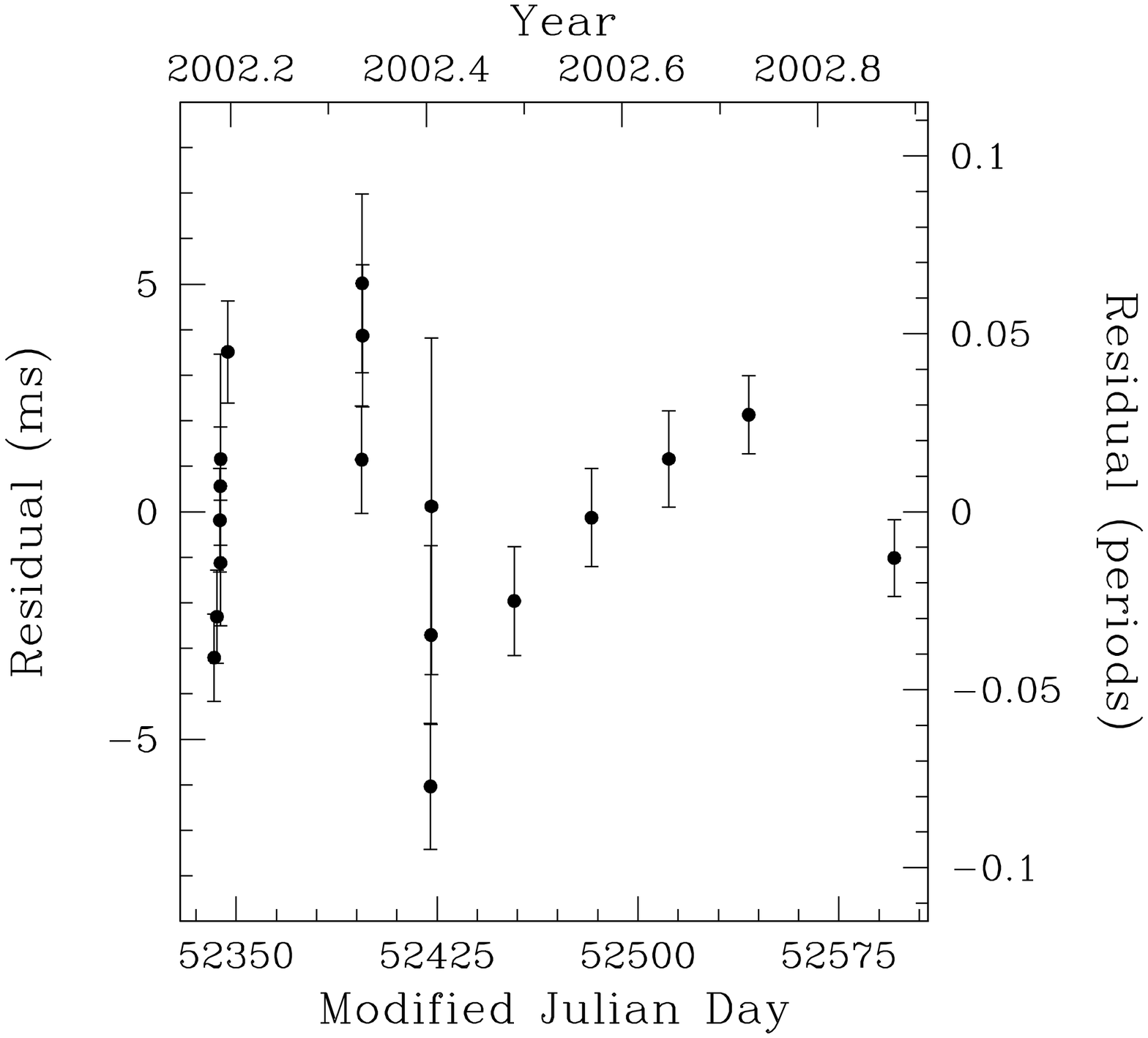}
\caption{Post-fit residuals for \psr\ after removal of the best-fit 
spin parameters in Table~\ref{ta:spin}. \label{fig:residuals}}
\end{center}
\end{minipage}
\end{figure}

In the timing analysis, we included only those events having energies in the 2--30~keV energy range so as to maximize the signal-to-noise ratio of the pulse.  Each binned time series was epoch-folded using the best estimate frequency determined initially from either a periodogram or Fourier transform (though later folding was done using the timing ephemeris determined by maintaining phase coherence; see below). Resulting pulse profiles were cross-correlated in the Fourier domain with a high signal-to-noise template created by adding phase-aligned profiles from previous observations. The 2--30~keV template is shown in Figure~\ref{fig:profile}.  The cross-correlation returns an average pulse time-of-arrival (TOA) for each observation corresponding to a fixed pulse phase. The pulse phase $\phi$ at any time $t$ can be expressed as a Taylor expansion,
\begin{equation}
\phi(t) = \phi(t_0) + \nu_0 (t-t_0) + \frac{1}{2} \dot{\nu}_0(t-t_0)^2   + \frac{1}{6} \ddot{\nu}_0(t-t_0)^3 + \cdots,
\label{eq:phase}
\end{equation}
where $\nu\equiv 1/P$ is the pulse frequency,  $\dot{\nu} \equiv d\nu/dt$, etc. and subscript `$0$' denotes a parameter evaluated at the reference epoch $t=t_0$. The TOAs were fit to the above polynomial using the pulsar timing software package \texttt{TEMPO}\footnote{http://pulsar.princeton.edu/tempo}. Unambiguous pulse numbering is made possible by obtaining monitoring observations spaced so that the best-fit model parameters have a small enough uncertainty to allow prediction of the phase of the next observation to within $\sim0.2$, this was accomplished by two closely spaced observations  (within a few of hours of each other) followed by one spaced a few days later. We are now capable of maintaining phase coherence with regular monitoring observations.

Arrival time residuals for \psr\ are shown in Figure~\ref{fig:residuals}. 
Spin parameters determined via phase-coherent timing 
using nine months of \xte\  data
are presented in Table~\ref{ta:spin}.  Phase residuals from the simple model
given in Table~\ref{ta:spin} are random, though not completely consistent
with zero given the uncertainties, which are typically $\sim$1~ms.  
This suggests there might be some pulse ``jitter,'' as is sometimes
seen in radio pulsar timing.
The tabulated spin parameters include a significant
frequency second derivative.  These parameters imply
a braking index $n = 1980 \pm 93$.  This number is
much larger than the canonical value of 3.  This could
reflect highly unusual spin-down behavior, unprecedented
in rotation-powered pulsars.  If so, the initial spin
period would not be determined for this pulsar.  
A more likely scenario is that the large value measure for $n$ 
reflects conventional timing noise, common to rotation-powered
pulsars.  In this case, a future measurement of the true, 
deterministic value of $n$ is not precluded.   

\subsection*{Spectroscopy}

This source is of particular interest given its relatively hard spectrum,
as, unlike most X-ray detected pulsars, pulsed emission well above
10~keV is definitely detected.
We studied the pulsed spectrum by summing all our folded, phase aligned  
time series into two  energy bands (2--10~keV and 10--30~keV). The 
folded profiles in these energy bands were identical. The 2--30~keV profile is 
displayed in Figure~\ref{fig:profile}.

\begin{table}
\begin{center}
\begin{tabular}{ll}
\hline\hline
Parameter & Value \\
\hline
$\Gamma$   & $1.78\pm 0.74$  \\
$n_H$  ($\times 10^{22}$\ cm$^{-2}$ ) &  2.36 (fixed)  \\
Flux  ($ \times 10^{-12}$ erg\ cm$^{-2}$\ s$^{-1}$) & $1.18\pm 0.36$   (2$-$10 keV) \\
$L_{\mathrm{x}}$ ($\times 10^{32}$ erg s$^{-1}$) & $2.81\pm 0.86$ (2$-$10 keV)\\
$\chi_{\nu}^2$ ($\nu$) &  0.88 (21)\\
\hline
\end{tabular}
\end{center}
\caption{Spectral fit of  the pulsed component of \psr\ to a simple photoelectrically absorbed power-law model of photon index $\Gamma$. The errors quoted indicate $90\%$ confidence intervals. The X-ray luminosity $L_x$, in the 2$-$30~keV range, was calculated assuming a distance of 5~kpc (Becker et al., 1985; Green, 1998) and assuming a beaming fraction of 1~sr. \label{ta:spectra}}
\end{table}

In order to obtain a high signal-to-noise pulsed spectrum for  \psr, 
data from each observing epoch were folded at the expected pulse period as was done for the timing analysis. However, here, 8 phase bins were used across the pulse. For each phase bin, we maintained a spectral resolution of 128 bins over the PCA range. One off-pulse phase bin was used as a background estimator. The pulse profiles were then phase aligned, so that the same off-pulse bin was used for background in every case. The remaining phase bins were summed, and their spectral bins regrouped into groups of 4 using the \texttt{FTOOL} \texttt{grppha}. Energies below 2~keV and above 70~keV were ignored, leaving 23 spectral channels for fitting. The regrouped, phase-summed data set, along with the background measurement, were used as input to the X-ray spectral fitting software package \texttt{XSPEC}\footnote{http://xspec.gsfc.nasa.gov}. Response matrices were created using the \texttt{FTOOL}s \texttt{xtefilt} and \texttt{pcarsp}. We fit the data using a simple photoelectrically absorbed power law, holding only $N_H$ fixed at the value found by \citet{rt+02} (see Table~\ref{ta:spectra}). 
 Uncertainties were  measured using the \texttt{XSPEC} command \texttt{steppar}. The photon index $\Gamma$ we obtained is consistent within the 
uncertainties with the value ($\Gamma = 1.11^{+0.37}_{-0.11}$) found by \citet{rt+02} using \textit{Chandra} 
observations. \nocite{bmd85,gre98}

\begin{figure}
\begin{center}
\includegraphics*[height=0.45\textwidth]{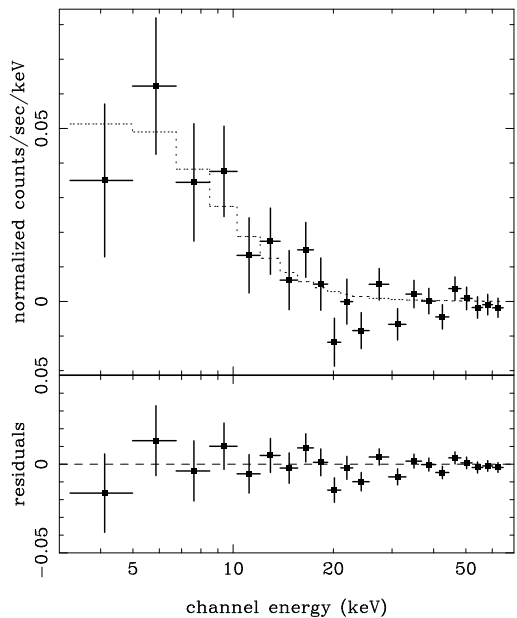}
\caption{Top: Phase-averaged spectrum of the pulsed component of \psr. The dash-line indicates a simple photoelectrically absorbed power-law model. Bottom: residuals after subtraction of the model. \label{fig:spectra}}
\end{center}
\end{figure}

\section*{DISCUSSION}

The ratio  of the pulsed X-ray luminosity in the 2--10~keV band to $\dot{E}$ gives an efficiency,
\begin{equation}
\eta_{\mathrm{x}} \equiv  \frac{L_{\mathrm{x}}}{4\pi I\dot{P}/P^3} = 0.06\%,
\end{equation}
assuming isotropic emission. Comparing  the efficiencies calculated  by
\citet{pccm02} for the 41 known rotation-powered X-ray pulsars, \psr\ is the third most efficient, with   PSR~J1846--0258 \citep{gvbt00} being the first, and PSR~B1509--58 \citep{cm+01} the second. Note that the efficiencies calculated by \citet{pccm02} often included  unpulsed contributions resulting from pulsar wind nebular emission, therefore their efficiencies can only be considered upper limits to the pulsed efficiencies.  
It is interesting to note the similarities among the three most X-ray efficient rotation-powered   pulsars: (i) they all have long duty-cycles, with profiles that remain broad even to high energies ($>10$~keV) (ii) they have very hard spectra with photon indices $\Gamma \sim 1.1 - 1.3$  (iii) they are all very young ($<2000$~yrs) pulsars in the centers of supernova remnants (iv) they all have pulsar wind nebulae.
Furthermore, none of these pulsars have been detected in  high energy $\gamma$-rays ($E>100$~MeV). This is in contrast to pulsars such as the Crab, Vela, and PSR~B1706--44 \citep{ghd02}, which have very narrow X-ray pulse profiles, relatively low X-ray efficiencies, but are all detected in high energy  $\gamma$-rays. 

The spectrum and pulse morphology can give us a clue to the emission mechanism at work in \psr. High-energy emission mechanism models generally fall into two  classes: polar cap models \cite[e.g.][]{hs+02} and outer gap models \cite[e.g.][]{rom02}. The two harder  components of the four component X-ray profile (in the \xte\ band)  of the Vela pulsar, which are as hard or  harder that \psr\,  are due to particle  curvature emission according to the polar cap model \cite[see][]{hs+02}. These  two components are 
 very narrow and phase aligned with the double  peaked $\gamma$-ray pulse. 
For \psr, the broad profile and the 
absence of $\gamma$-ray pulsations make a similar mechanism unlikely. In the outer gap model, the broad, hard profile of \psr\ would be  from the tail of the synchrotron component. According to \citet{rom02},
the peak of the synchrotron spectrum for \psr\ would occur at $E\sim 31$~MeV, this could be verified  with  $\gamma$-ray observatories  such as \textit{GLAST} and \textit{INTEGRAL}.

\section*{ACKNOWLEDGMENTS}

This work was supported in part by
a NASA LTSA grant, an NSERC Research Grant.
VMK is a Canada Research Chair.  This
research has made use of data obtained through the High Energy
Astrophysics Science Archive Research Center Online Service, provided
by the NASA/Goddard Space Flight Center.

\bibliographystyle{apalike}
%\bibliography{journals1,modrefs,psrrefs,crossrefs}

\vspace*{2\baselineskip}
\noindent E-mail address of V.M. Kaspi \ \ \ \url{vkaspi@physics.mcgill.ca}

\end{document}